\documentclass[12pt]{article}
\usepackage{graphicx}

\begin{document}

\title{A note and a new perspective on particle horizon problem}
\author{ Zhang Hongsheng\\zhanghs@itp.ac.cn\\
Institute of theoretical physics, Chinese academy of sciences \\ Beijing,
100081, P. R. China}
\date{\today}
\maketitle
% ------------------------------------------------------------------------

%%% ----------------------------------------------------------------------
\begin{abstract}
   We cautiously reanalyze some easily confused notions on particle horizon problem in this paper and
   then we give a new answer to the particle horizon problem. This answer is independent of physics
   plunging into Planck time.

\end{abstract}

%000000000000000000000000000000000000000000000000000000000000001
\section{Introduction}
  Particle horizon problem is an interesting problem in cosmology. It origins in big bang model, where
  people find that if we go back to the early universe we can see the observed universe is divided into
  many causal unconnected regions, which need people answer the question--why the observed universe is
  homogeneous? The inflation model seems like a good candidate to the question[1,2]. In fact inflation
  is only a project but not a specific model. Specifically we choose the Guth's "old inflation model"[1].
   The methods of analyzing on particle horizon problem are almost the same to all inflation models.
   The literatures usually did not analyze particle horizon problem directly. Some mistakes or
   misunderstandings then arise, for example the figure of particle horizon Guth and Steinhardt giving  
   in a popular science magazine[3]. Many authors think "the perturbations exit horizon(they mean Hubble radius)
   have no casual relation", for example[4]. We will point their mistakes and give the correct answers.
   Also do we point out the weakness of the inflationary solution to particle horizon problem.

   In this paper we will give a method to solve particle horizon problem without inflation.

   We always discuss in 4-dimensional universe and the density is critical density.

%00000000000000000000000000000000000000000000000000000000000000000002
\section {Particle horizon and Hubble radius }
\subsection{On matter dominated epoch and radiation dominated epoch}
  In FRW(Friedmann-Robertson-Walker) universe whose density is critical density the metric can be written as

\begin{equation}
ds^2=dt^2-a^2(t)(dx^2+dy^2+dz^2),
\end{equation}

where $a(t)$ is scale factor.
In matter dominated epoch,

\begin{equation}
a(t)=c_1t^{\frac{2}{3}},
\end{equation}

where $c_1$ is a constant, which leads the particle horizon

\begin{equation}
H_{ph}=a(t_{ph})\int^{t_{ph}}_0\frac{1}{a(t)}dt=3t_{ph},
\end{equation}

and the Hubble radius

\begin{equation}
H_r=\frac{1}{H(t_r)}=\frac{a(t_r)}{\frac{da}{dt}|_{t=t_r}}
=\frac{3t_r}{2}.
\end{equation}

  It is easy to see on the same simultaneous hypersurface of standard 1+3 decomposition of FRW universe
  where $t_{ph}=t_r$ the particle horizon and the Hubble radius have the relation

\begin{equation}
H_{ph}=2H_r.
\end{equation}

Through almost the same way we obtain the relation of the particle horizon and the Hubble radius in the epoch of
 radiation dominated as follow

\begin{equation}
H_{ph}=H_r=2t_{ph}=2t_r.
\end{equation}

  So we can say in radiation or matter dominated epoch the Hubble radius can crudely represents the particle horizon.

\subsection{On vacuum dominated epoch}

In most inflation models there is an epoch when the vacuum
dominates. In that epoch the spacetime metric is de Sitter metric

\begin{equation}
ds^2=dt^2-a^2(t)(dx^2+dy^2+dz^2),
\end{equation}

where

\begin{equation}
\begin{array}{l}
a(t_0)=a_0, \\
a(t)=a_0e^{\chi(t-t_0)}.
\end{array}
\end{equation}

 $\chi$ is a parameter, $t_0$ denotes the time when the universe enters vacuum dominated epoch and $t$
 denotes any time in the vacuum dominated epoch.
  The particle horizon in the vacuum dominated epoch can be calculated as follow

$$H_{ph}=a(t_{ph})\int_{t_0}^{t_{ph}}\frac{1}{a(t)}dt$$

\begin{equation}
=H_{ph}(t_0)+\frac{1}{\chi}(e^{\chi (t-t_0)}-1),
\end{equation}

  Mimicking the definition of Hubble radius in matter and radiation dominated epoch we define Hubble radius

\begin{equation}
H_r=\frac{1}{H}=\frac{1}{\chi}=constant.
\end{equation}

It is obvious the Hubble radius $can not$ represent the particle
horizon in vacuum dominated epoch. So the sentences such as the
perturbation waves exited Hubble horizon(means Hubble radius) have
no casual relation is nonsense.

\section{Particle horizon problem in inflation model}
  It is generally believed that the particle horizon problem in big bang model has been conquered by
  inflation model. But this result need carefully consider. Let's review the horizon problem.  As approximate
  calculation we take present Hubble constant $H_0=65km.sec^{-1}.Mpc^{-1}$ and the universe is at critical
  density. In big bang model from the first section the particle horizon is

\begin{equation}
H_{ph}=2H_r=\frac{2}{H}=2.85*10^{26}m.
\end{equation}

  The presently observable universe is the section of the whole universe in particle horizon now. Based on
  the red-shift of galaxy and the red-shift of CMB(cosmological microwave background) we know the observed
  universe is the same order of the presently observable universe, which means the supposition that most
  time of our universe is dominated by matter is correct. So in practical number calculation we consider
  the presently observable universe is the same as the observed universe. We have known the fact that the
  observed universe is homogenous. Let's consider the presently observed universe and particle horizon when
   we come back to the early universe. We know the radius of the observed universe $D_b$ is proportional
   to $a(t)(=c_1t^{\frac{2}{3}}$,for matter dominated;$=c_2t^{\frac{1}{2}}$,for radiation dominated) where
   the radius of the particle horizon $D_{ph}$ is proportional to $t(=3t$,for matter dominated;$=2t$ for
   radiation dominated) and they are of the same order now. So when we come back to the early universe we
   find that the universe we observed now is perfectly out of horizon. For example when $t=10^5$ year

\begin{equation}
D_b(t)=D_b(t_n)* (\frac{t}{t_n})^{\frac{2}{3}}=1.3*10^{23}m,
\end{equation}

where $t_n=\frac{2}{3}\frac{1}{H}$ is the age of our universe. At the same time the particle horizon is

\begin{equation}
D_{ph}(t)=D_{ph}(t_n)* \frac{t}{t_n}=2.8*10^{21}m.
\end{equation}

  We can see that $D_b=46*D_{ph}$ when $t=10^5$ year. The problem is more serious when we come back nearer to
  the big bang singularity. In the very early universe the universe is radiation dominated, but the conclusion
  is the same. When we trace back to the Planck epoch $t_p=10^{-43}s$ according to standard big bang model,
  following the method we just mentioned we can obtain $D_b(t_p)=10^{30}D_{ph}(t_p)$. If we require the universe
   we observed now was in particle horizon one time, there must exist an epoch when the scale factor $a(t)$
   grows $10^{30}$ times than the radius of particle horizon.
  Let's take a look on inflation model. There are many kinds of inflation models. The core of inflation model
  is the universe has undergone a much faster increasing than the standard FRW model. The inflation is vacuum
   dominated epoch. Supposing the initial time of inflation is $t_i$ and the ending time is $t_e$. Also will
   we point that in most literatures people use the "Hubble radius" represents the particle horizon. From the
   second section we see that in matter dominated epoch or radiation dominated epoch it is right.  The key
   flaw of past analysis, such as [4], are people mistakenly regard the Hubble radius as the particle horizon
   in the vacuum dominated epoch. Based on the second section the correct analysis of particle horizon in
   vacuum dominated epoch is presented here. Generally we suppose before inflation the universe is dominated
   by radiation.  From Eq.(9)

\begin{equation}
H_{ph}=H_{ph}(t_0)+\frac{1}{\chi}(e^{\chi (t-t_0)}-1).
\end{equation}

  From Eq.(6)

\begin{equation}
H_{ph}(t_0)=2t_0,
\end{equation}

  Define enlargement factor in inflation epoch

\begin{equation}
Z=\frac{x_{t_e}}{x_{t_i}},
\end{equation}

where $t_i$ is initial time of inflation , $t_e$ is the ending time of inflation and $x$ is arbitrary function.

  The enlargement factor of scale factor $a(t)$ is

\begin{equation}
Z_a=\frac{a(t_e)}{a(t_i)},
\end{equation}

from Eq.(8)

\begin{equation}
Z_a=e^{\chi(t_e-t_i)}.
\end{equation}

The enlargement factor of particle horizon $H_{ph}$ is,from Eq.(14),

\begin{equation}
\begin{array}{l}
Z_{H_ph}=\frac{H_{ph}(t_e)}{H_{ph}(t_i)}
        =\frac{1}{2} e^{\chi(t_e-t_i)}+\frac{1}{2}.
\end{array}
\end{equation}

  The parameter $\chi$ can be written as[1]

\begin{equation}
\chi={(8\pi G \rho/3)}^{(1/2)}=10^{34}s^{-1},
\end {equation}

where $\rho$ is chosen as GUT density scale[1].
 The proportion of enlargement factor of $a$ to the enlargement factor of $H_{ph}$ can be written as

\begin{equation}
\frac{Z_a}{Z_{ph}}=2.
\end{equation}

 Let $t_i=10^{-34}s$ and $t_e=10^{-32}s$, we can calculate both $ Z_a$ and $Z_{ph}$ are of order $10^{43}$.

\section{A thorough analyzing on particle horizon problem in inflation model}

Without inflation, we know the size of observed universe is
$10^{30}$ times to size of casual connected region when we trace
back to Planck time. If we think the size of observed universe to
be the $same$ size of the presently observable universe, or
equivalently the present particle horizon, which requires there
exists an epoch when the proportion of enlargement factor of $a$
to the enlargement factor of $H_{ph}$ is $10^{30}$, we see the
inflation model can't do its work. We need restudy the observed
universe and the observable universe counting inflation. The
contents of observed universe ,such as galaxies and CMB, are all
formed quite after inflation. The distances of far celestial
bodies are  measured with methods not affected by physical
procession before the formation of these bodies. So the size of
observed universe in big bang model is as the same as the size of
observed universe in inflation model.  But the age of the universe
and the observable universe are different. They rely on the whole
procession of the universe, from big bang to now. So it is
necessary to investigate the influence of inserted inflation on
the them. From the following proofs we can see the inflation
hardly do any work on the age of the universe and may do some work
on the observable universe.

Let us consider the solution of Friedmann equation with arbitrary constants on condition that matter
dominates. Because the properties of radiation dominated epoch is much alike the the properties of
radiation dominated epoch comparing to the properties of vacuum dominated epoch. we may use the equations
for matter dominated to substitute equations for radiation dominated for estimation.

\begin{equation}
D_b(t)=(B_1t+B_2)^{2/3},
\end{equation}
where $D_b(t)$ is the size of presently observed universe at time $t$,$ B_1 $ and $B_2$ are arbitrary
constants decided by boundary conditions.

\begin{equation}
D_b(t_n)=(B_1t_n+B_2)^{2/3},
\end{equation}

\begin{equation}
D_b(t_e)=(B_1t_e+B_2)^{2/3},
\end{equation}

where $t_n$ is the time from big bang to now (considering inflation), $t_e$ is the time when
inflation ends. Here we think the matter dominated epoch begin at $t_e$ for estimation. If we
exclude inflation ,the equation can be written as, roughly thinking matter dominating through
all universe history,

\begin{equation}
D_b(t)=D_b(t_n)(\frac{t}{t_n})^{2/3}.
\end{equation}

It is easy to obtain

\begin{equation}
t_n=\frac{2}{3}\frac{\frac{dD_b}{dt}}{D_b}=\frac{2}{3}H^{-1}(now),
\end{equation}

It is obviously that only if we can proof that
$\frac{\frac{B_2}{B_1}}{t_n}$ is a small quantity we proof the
inflation's influence on the age (which is calculated by Hubble
constant) can be literally omitted. In fact, from Eq.(23) and
Eq.(24)

\begin{equation}
\frac{B_1t_n+B_2}{B_1t_e+B_2}=(z+1)^{3/2},
\end{equation}

then

\begin{equation}
(z+1)^{3/2}-1=\frac{t_n-t_e}{t_e+\frac{B_2}{B_1}}<\frac{t_n}{t_e+\frac{B_2}{B_1}}<\frac{t_n}{\frac{B_2}{B_1}},
\end{equation}

\begin{figure}
\centering
\includegraphics[totalheight=3.5in]{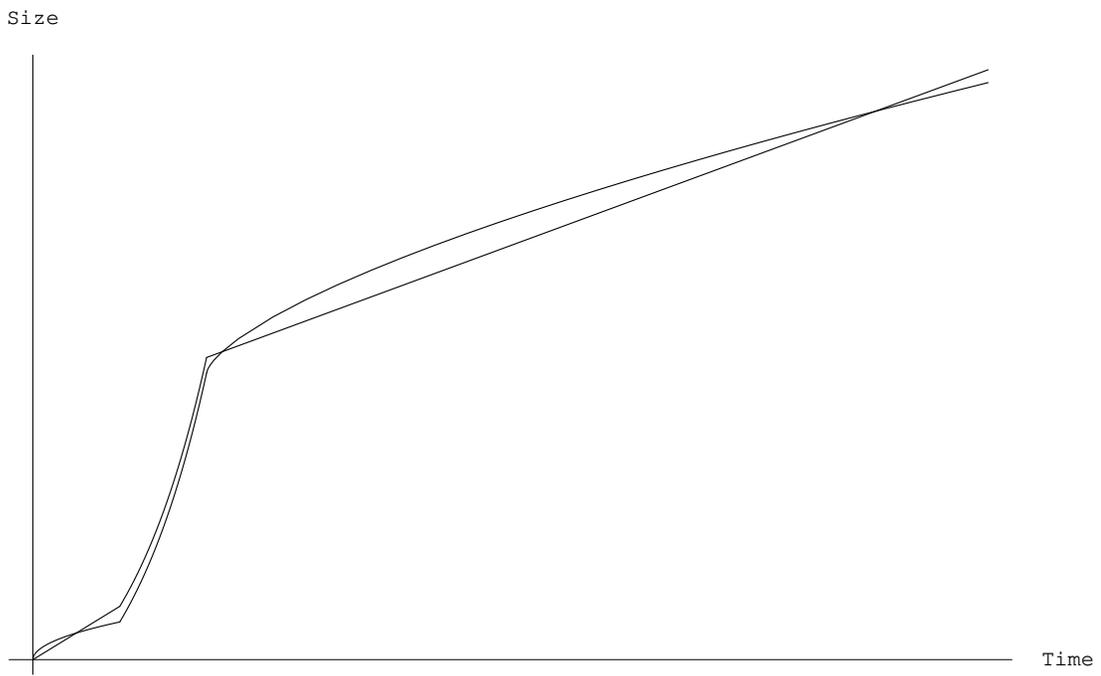}
\caption{The size of particle horizon vs the size of observed
universe } \label{fig1}
\end{figure}

where $z$ is redshift factor, which is much larger than redshift factor of CMB(1100), because the end of
inflation is much earlier than the CMB decoupling. So the age of our universe can also be denoted by Hubble
 constant in inflation model. To the particle horizon thing is different. Based on the Section 2 the calculation
 is straightforward. The result is plotted on the figure 1, which is just a sketch. The curve which is upper at
 the end of time represents particle horizon and the other represents presently observed universe. We give a
 summary on the particle horizon problem and relative problems in inflation model comparing to big bang.\\

1. The age of universe is shortened, but little, which means if inflation really happened the true age
of our universe is shorter than we think in big bang model(experimentally measured by $H^{-1}$). \\

2. The presently causal connected region (or the particle horizon
,or the the presently observable universe) is larger ,more or
less,depending on the enlargement factor of
$\frac{a(t_e)}{a(t_i)}$, than we think in big
bang model.\\

3. The observed universe may intersect the particle horizon one, two or three times depending on
 $\frac{a(t_e)}{a(t_i)}$.\\

It is clear the particle horizon is larger than the observed
universe(larger by about $6*10^{17}m$). The two curves intersect
three times. Simple calculation give
them$(t_1=2.2*10^{-83}s,t_2=3.06*10^4s,t_3=3.16*10^{17}s)$.
 Just the same calculation gives $6*10^{24}m,t_1=2.2*10^{-97}s,t_2=1.25*10^{15}s,t_3=2.95*10^{17}s$ in the situation
 of Guth and Steinhardt[3]($\frac{a(t_e)}{a(t_i)}=10^{50},t_i=10^{-34}s$), which is very different from
 their figure where no intersect is presented after inflation. Note that the first intersect point in both
 of them make no sense because of immersing into Planck epoch. Obviously if $\frac{a(t_e)}{a(t_i)}$ is large
 enough the situation of no intersect after inflation appears. The critical point is
 $Z_c=\frac{a(t_e)}{a(t_i)}=7.02*10^{50}.$ If the enlargement factor $Z<Z_c$ there exist
 bulks(parts of observed universe) exit and then reenter the particle horizon after inflation.
  But this is fundamentally deferent from reference[4]--there are no particle horizon exiting in
   inflation epoch in our scenario. So a matter bulk crossing apparent horizon(Hubble radius) during
   inflation does not mean its outer parts losing causal connection.
  We can see this scheme for particle horizon problem depends on the physics before Planck time,
  which we are not sure. It has to suppose the equation $H_{ph}=2t$ from $t=0$. In some sense we
  do not know if we can say "before" when physics plunging into Planck time. The same problem exists
  in the analysis of big bang model without inflation. In next section we will give a new project to
  conquer the particle horizon problem. This project is independent of physics plunging into Planck time.

\section{Solve the particle horizon problem}
   The particle horizon is relative to a particle(an observer). Holding this in mind tightly we solve the
   particle horizon problem only need an idea.\\
   In figure 2 A,B,C are three points and the circles around them denote the particle horizons separately at
   time $t=t_0$. Point 1 and 2 in the particle horizon of A, 2 and 3 in B, etc. We can choose an physical
   quantity, such as temperature $T$ to represent the thermal equilibrium. The early literature[4] has
   proofed the very early universe can reach thermal equilibrium in the particle horizon of a particle,
    such as A,B,C, which equal to say

\begin{equation}
T_1=T_2, T_2=T_3,T_3=T_4.
\end{equation}

\begin{equation}
\Rightarrow T_1=T_4,
\end{equation}

\begin{figure}
\centering
\includegraphics[totalheight=2in]{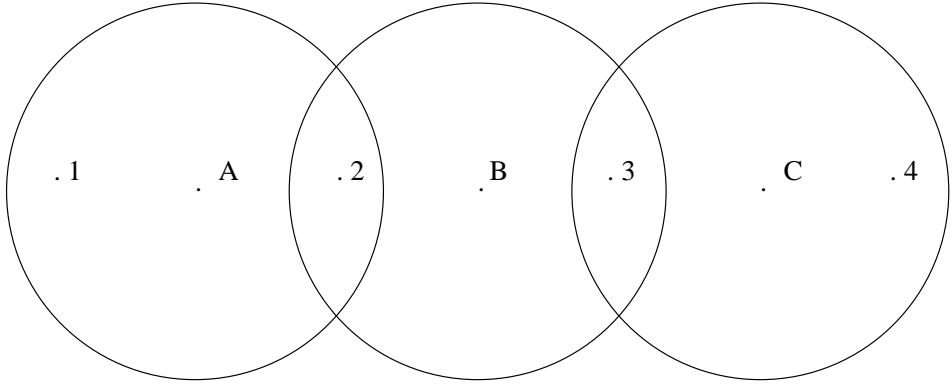}
\caption{Particle horizons }
\label{fig2}
\end{figure}

although the two points have no casual relation. Processing this
derivation again and again, we get the conclusion that all the
universe despite whether there exists casual relation in it is in
thermal equilibrium. The horizon of an arbitrary point encloses a
neighbor of the point. A compact region of a spacetime manifold
can be covered by limited such open sets(neighbors of spacetime
points).   This proof seems reasonable. But if we check it enough
cautiously we can find a flaw. The flaw is "the circles around
them denote the particle horizons separately at time $t=t_0$." It
is no problem we can set up local RW coordinates in a particle
horizon because of the homogeneous of spacetime in a particle
horizon. But generally speaking no such good coordinates exist on
the manifold (We always consider physics after the universe can be
regarded as a manifold). So we must show particle horizons of
different points share the same simultaneous hypersurface. In
figure 3, $l$ and $m$ are two world lines representing two
particles in communal parts of two particle horizons, say, of A
and B in figure 2. Supposing the two particle horizons have
different simultaneous hypersurfaces $DE$ and $DF$ we will get
contradictory result. We choose a physical quantity representing
physical homogeneousness, for example, temperature $T$. Because of
the thermal equilibrium in a particle horizon in the very early
universe, we have

\begin{figure}
\centering
\includegraphics[totalheight=3in]{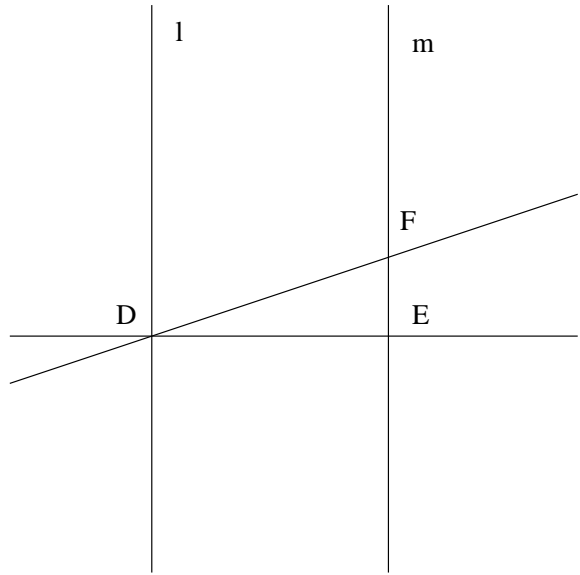}
\caption{Simultaneous hypersurfaces } \label{fig3}
\end{figure}

\begin{equation}
T_D=T_E,
T_D=T_F,
\end{equation}

\begin{equation}
\Rightarrow T_F=T_E,
\end{equation}
which contradicts to the fact that the universe is not static. So we conclude that the whole universe
 manifold share a communal slicing if we require the space is homogeneous in a particle horizon and the
 universe is not static, which gives the proof of Eq.(29) and Eq.(30) a sound base.\\

A puzzle may appear. We know no particle travel faster than light
in vacuum. How do the causal discrete areas "know" temperature of
each other? We give a naive analogy of this situation. A theater
has fifty rows. All of the rows are full except the first row. Now
all the audience stand up and then sit down to the front row,
which seems like the people in the last row reach to the first raw
at once.

%000000000000000000000000000000000000000000000000000000000000000000000005
\section*{Acknowledgments}
This research is  supported by a grant from Chinese academy of sciences.

% 00000000000000000000000000000000000000000000000000000000000000000000000


\begin{thebibliography}{99}
\bibitem {1} Guth,A.H.,Phys. Rev.,D23,347-356.
\bibitem{2} Albrecht,A. and Steinhardt,P.J. Phys. Rev. Lett.,48,1220-1223.

\bibitem{3} Guth,A.H. and Steinhardt,P.J. Scientific American,May, 1984,Vol 250,90-102.
\bibitem{4} Kolb,E.W. and Turner,M.S.,The early universe,Addison-Wesley publishing company,1990.
\end{thebibliography}
\end{document}